\documentstyle[prb,aps,epsf,rotate,amstex,amssymb]{revtex}
\begin{document}
\draft
\makeatletter
\title{Field distributions and effective-medium approximation 
for weakly nonlinear media}
\author{Yves-Patrick Pellegrini\thanks{e-mail: {\protect\tt 
 pellegri@bruyeres.cea.fr}}}
\address{ Service de Physique de la Mati\`ere Condens\'ee, \\
		Commissariat \`a l'Energie Atomique,\\
		BP12, 91680 Bruy\`eres-le-Ch\^atel, France.}
\date{Last modified: November 18, 1999. Printed: \today}
\maketitle
\makeatother
\begin{abstract}
An effective-medium theory is proposed for random weakly nonlinear dielectric 
media. It is based on a new gaussian approximation for the probability 
distributions of the electric field in each component of a multi-phase 
composite. These distributions are computed to linear order from a 
Bruggeman-like self-consistent formula. The resulting effective-medium formula 
for the nonlinear medium reduces to Bruggeman's in the linear case. It is exact 
up to second order in a weak-disorder expansion, and close to the exact result 
in the dilute limit (in particular, it is exact for $d=1$ and $d=\infty$). In a 
high contrast situation, the noise exponents are $\kappa=\kappa'=0$ near the 
percolation threshold. Numerical results are provided for different 
weak nonlinearities. The use of the Bruggeman formula as a starting point 
for nonlinear homogeneization theories in dimensions $d>2$ is questioned on 
the basis of known exact bounds on the noise exponents.
\end{abstract}
\bigskip

\makeatother
\section{Introduction}
\label{intro}
Numerous papers have been devoted to the problem of determining the effective 
constitutive law of random nonlinear composites in electrostatics 
\cite{AHAR87,STRO88,ZENG88,BERG89,BLUM89,KOTH90,PONT92,GU92,YU93,YU94,LEE95,GAO96,WAN96,SALI97,PONT97,BART98}, 
of importance for understanding 
various nonlinear phenomena, either entering the class 
of ``weakly nonlinear'' (WNL) phenomena, or that of ``strongly nonlinear'' 
phenomena such as random fuse-type or dielectric breakdown-type failure of 
materials \cite{DUXB86}. 

A convenient approach to the homogeneization problem is the ``energetic'' 
one \cite{HILL63,PONT92b}, which can be summarized as follows. Let $w_x(E)$ 
be the local potential (the energy density) from which derives the local 
nonlinear constitutive law. The homogeneization step consists in computing 
the effective potential $W$ as the volume average
\begin{equation}
W=\langle w_x(E)\rangle,
\end{equation}
from which the effective constitutive law can be deduced (see below).
Denoting by $p_\alpha$ the volume proportion of component (``phase'') 
$\alpha$ in a composite medium where $w_x=w_\alpha$ if $x$ is in phase 
$\alpha$, and introducing the volume averages {\em per phase} 
$\langle\cdot\rangle_\alpha$, $W$ can be written
\begin{equation}
\label{weff}
W=\sum_\alpha p_\alpha \langle w_\alpha(E)\rangle_\alpha.
\end{equation}

A class of effective-medium theories, mostly concerned with weak 
power-law nonlinearities, use our ability to compute the second moment 
of the electric field $E$ in the various phases of the composite 
\cite{BERG78}, completed by a decoupling approximation \cite{ZENG88,HUI95a}
\begin{equation}
\label{decoupl}
\langle w_\alpha(E)\rangle_\alpha\simeq 
w_\alpha\bigr(\langle E^2\rangle_\alpha^{1/2}\bigl).
\end{equation}
This approximation is exact only when $w_\alpha$ is quadratic 
(the linear theory), or when the field is uniform, which occurs only 
if all the $w_\alpha$ are equal, i.e.\ in a homogeneous medium.

The purpose of this paper is to present a means to overcome this 
approximation, starting from the following remark: actually, with 
an additional hypothesis of ergodicity {\em per phase}, a volume phase 
average can be reinterpreted as an average over some probability 
distribution for the electric field in the phase under consideration. 
In this paper, volume averages  will henceforth be identified to 
statistical averages. Then, the problem of computing $W$, 
Equ.\ (\ref{weff}), reduces to computing the probability distribution 
${\cal P}_\alpha$ of the field in each phase, by means of which the 
phase averages $\langle\cdot\rangle_\alpha$ can be carried out, 
without appealing to (\ref{decoupl}). 

Note that the full probability distribution ${\cal P}({\mathbf E})$ 
of the electric field in the medium is then
\begin{equation}
\label{multimod}
{\cal P}({\mathbf E})=\sum_\alpha p_\alpha {\cal P}_\alpha({\mathbf E}).
\end{equation}
Thus, the probability distributions of the field {\em per phase} are 
functions of great importance, since they carry all the information 
needed to build up the effective nonlinear properties. A few analyses 
of ${\cal P}({\mathbf E})$ are available \cite{BART97}, mainly focused 
on its tail of relevance for breakdown phenomena \cite{DUXB95}. However, 
up to my knowledge, none has been specifically aimed at the individual 
components ${\cal P}_\alpha({\mathbf E})$. Chen and Sheng studied 
numerically the distribution of the {\em local} field in binary 
composites \cite{CHEN91}. An analytical model of the same distribution 
was subsequently proposed by Barth\'el\'emy and Orland, with excellent 
agreement with the latter simulations \cite{BART97}. Both works were 
limited to linear composites, in a weak-contrast situation. A first 
numerical attempt in the continuum, in a strong-contrast situation 
this time, is due to Cheng and Torquato \cite{CHEN97}. They carried 
out two-dimensional numerical calculations on disordered systems with 
inclusions of various shapes. An exact calculation for ${\cal P}$ in 
a {\em linear} medium with Hashin-Shtrikman (HS) geometry was recently 
made using a density of states approach \cite{CULE98}. In these 
works, the distribution ${\cal P}$ was found to be essentially 
bimodal for a binary medium (save in the HS geometry where 
fine-structure peaks are not wiped out by positional disorder), 
especially at high dielectric contrast \cite{CHEN97}. This is 
understandable in wiew of the decomposition (\ref{multimod}).

Still, in parallel with the second moment of the field, the 
{\em first} moment in each phase (the average field) can be computed 
as well from any linear effective-medium theory, without requiring 
any {\it a priori} knowledge about ${\cal P}_\alpha$. Widely used in 
mechanics of random continuous media in the framework of the 
so-called ``thermoelastic problem'' \cite{WILL81}, this fact seems 
to have escaped most of the litterature on random dielectric media, 
excepted in the work by Ponte Casta\~neda and Kailasam \cite{PONT97} 
where it is used to compute reference fields around which non-linear
 potentials are expanded prior to homogeneization. Knowing the first 
two moments is enough to attempt a modelization of ${\cal P}_\alpha$. 
The proposal examined in the present paper is to approximate each 
${\cal P}_\alpha({\mathbf E})$ by a gaussian {\em vector} 
distribution, and thereby, to estimate (\ref{weff}) for any potential, 
without appealing to the strong (and somewhat artificial) decoupling 
assumption (\ref{decoupl}). 

The scope of the paper is limited however to WNL media, for which it 
is sufficient to compute the field distribution to linear order only 
\cite{AHAR87,STRO88}. An extension of these considerations to 
strongly nonlinear media will be examined elsewhere.  

\section{General framework}
\label{th}
Let us therefore consider a $d$-dimensional random medium described 
by WNL thermodynamic potentials of the type
\begin{equation}
\label{orig}
w_x(E)=w_x^{\text{lin}}(E) +w_x^{\text{nl}}(E),
\end{equation}
where 
\begin{equation}
w_x^{\text{lin}}(E)=\frac{1}{2} \varepsilon E^2
\end{equation}
($\varepsilon(x)$ being the local permittivity entering the linear part), 
and $w_x^{\text{nl}}(E)$ is the nonlinear (i.e.\ non-quadratic) part of 
the potential, assumed to be small. The local constitutive law is
\begin{equation}
D_i=\frac{\partial w_x(E)}{\partial E_i}. 
\end{equation}

The potential is prescribed in each phase: $w_x(E)=w_\alpha(E)$ if $x\in\alpha$. 
Under the ergodic hypothesis introduced above, and a ``boundary'' condition    
\cite{NOTE1} $\langle {\mathbf{E}}\rangle={\mathbf{E}}^0$, the effective 
potential reads
\begin{equation}
W(E^0)=W_{\text{lin}}(E^0)+W_{\text{nl}}(E^0),
\end{equation}
where
\begin{eqnarray}
\label{lin}
W_{\text{lin}}(E^0)&=&\frac{1}{2}\sum_\alpha p_\alpha \varepsilon_\alpha 
\langle E^2\rangle_\alpha,\\
\label{nonlin}
W_{\text{nl}}(E^0)&=&\sum_\alpha p_\alpha \langle w_\alpha^{\text{nl}}(E)
\rangle_\alpha.
\end{eqnarray}  
The effective constitutive law in the homgeneized medium is then
\begin{equation}
\label{constithom}
D^0_i={\partial W(E^0)\over \partial E^0_i}.
\end{equation}

To first order in $w_x^{\text{nl}}$, the field distributions entering 
(\ref{lin}) and (\ref{nonlin}) need only be computed at the linear 
level \cite{AHAR87,STRO88}. The procedure is the following one. An 
extended linear {\em anisotropic} homogeneization problem is first 
considered, with an artificial ferroelectric-like part added to the 
potential \cite{PONT97}:
\begin{equation}
\label{ferroel}
\tilde{w}^{\text{lin}}_x(E)=\frac{1}{2} E_i\varepsilon_{ij} E_j+ 
P_{i} E_i, 
\end{equation}
where $\varepsilon_{ij}(x)=\varepsilon^\alpha_{ij}$ and $P_i(x)=
P_i^\alpha$ if $x\in\alpha$. The corresponding local constitutive 
law is $D_i=\varepsilon_{ij} E_j+P_i$. Assuming one knows an estimate 
for
\begin{equation}
\tilde{W}_{\text{lin}}(E^0)=\langle \tilde{w}^{\text{lin}}_x(E)\rangle,
\end{equation}
(cf.\ below), the following {\em exact} relations hold for the distribution 
${\cal P}_\alpha$ of the original linear problem (\ref{lin}):
\begin{mathletters}
\label{avfield}
\begin{eqnarray}
\label{ave}
\left\langle E_i\right\rangle_{\alpha}&=&\frac{1}{p^\alpha}\left. 
\frac{\partial \tilde{W}_{\text{lin}}}{\partial P^\alpha_i}
\right|_{\varepsilon_{ij}=\varepsilon \delta_{ij}\atop P_i=0},\\ 
\label{avee}
\left\langle E_i E_j\right\rangle_{\alpha}&=&\frac{2}{p^\alpha}
\left.\frac{\partial \tilde{W}_{\text{lin}}}{\partial
\varepsilon^\alpha_{ij}}\right|_{\varepsilon_{ij}=\varepsilon 
\delta_{ij}\atop P_i=0}.
\end{eqnarray}
\end{mathletters}
I emphasize that the derivatives are computed at the point $P_i=0$, 
$\varepsilon^\alpha_{ij}=\varepsilon^\alpha \delta_{ij}$ where 
$\varepsilon^\alpha$ is the scalar permittivity in (\ref{orig}) 
and (\ref{lin}). There, $\tilde{w}^{\text{lin}}_x=w^{\text{lin}}_x$ 
and $\tilde{W}_{\text{lin}}=W_{\text{lin}}$. Note that (\ref{avee}) 
is a slight generalization of the formula \cite{BERG78}
\begin{equation}
\label{derivscal}
\langle E^2\rangle_\alpha=\frac{2}{p^\alpha}\frac{\partial 
W_{\text{lin}}}{\partial\varepsilon^\alpha}.
\end{equation}
Both (\ref{ave}) and (\ref{avee}) are particular cases of a more 
general theorem, the demonstration of which can be found in appendix 
B of the review by Ponte Casta\~neda and Suquet on nonlinear 
composites\cite{PONT98}. This theorem is a consequence of the 
equality $\langle D_i E_i\rangle=\langle D_i\rangle \langle E_i\rangle$ 
which holds only for particular boundary conditions, or in the 
infinite-volume limit if there exists an effective permittivity.  

Let us set
\begin{eqnarray}
M_i^\alpha&=&\left\langle E_i\right\rangle_{\alpha},\\
C^\alpha_{ij}&=&\left\langle E_i E_j\right\rangle_{\alpha}-
M_i^\alpha M_j^\alpha,
\end{eqnarray}
with $M^\alpha_i\equiv M^\alpha \hat{E}^0_i$ and $C^\alpha_{ij}
\equiv C^\alpha_\parallel \hat{E}^0_i\hat{E}^0_j+
C^\alpha_\perp(\delta_{ij}-\hat{E}^0_i\hat{E}^0_j)$, since the 
averages only depend on the direction $\hat{E}^0_i=E^0_i/E^0$ 
of the macroscopic field. I introduce now the gaussian approximation 
for the probability distributions ${\cal P}_\alpha$
\begin{equation}
\label{pae}
{\cal P}_\alpha({\bf E})={1\over \left[(2\pi)^d \det(C^\alpha)
\right]^{1/2}} e^{-{1\over 2}(E_i-M^\alpha_i)
{C^\alpha}^{-1}_{ij}(E_j-M^\alpha_j)},
\end{equation}
which constitutes the main ingredient of the theory. 

There remains to compute an approximation for the quantities 
$\left\langle E_i\right\rangle_{\alpha}$ and $\left\langle 
E_i E_j\right\rangle_{\alpha}$, in oprder to completely define 
the above gaussian distributions. Let us therefore estimate 
$\tilde{W}_{\text{lin}}(E^0)$ through an extension of the 
Bruggeman self-consistent effective-medium approximation\cite{BRUG35} 
to potentials of the form (\ref{ferroel}). The formula is 
given and discussed in Appendix A. Using the scalar 
permittivities and $P_i=0$, this estimate reduces to
\begin{equation}
W_{\text{lin}}(E^0)=\frac{1}{2}\varepsilon_0 {E^0}^2,
\end{equation}
where $\varepsilon_0$ is the usual Bruggeman effective 
permittivity\cite{BRUG35}. In addition, the derivatives 
(\ref{ave}) and (\ref{avee}) of the estimate for 
$\tilde{W}_{\text{lin}}(E^0)$ yield
\begin{eqnarray}
M^\alpha&=&\mu^\alpha E^0,\\
\label{ee}
\left\langle E_i E_j\right\rangle_\alpha&=&z
{\mu^\alpha}^2\Biggl[E^0_iE^0_j
-\frac{{E^0}^2}{d+2}\left(1-y\langle\mu^2
\rangle\right)\delta_{ij}\Biggr],
\end{eqnarray}
where
\begin{eqnarray}
\mu^\alpha&=&\frac{d\varepsilon^0}
{\varepsilon^\alpha+(d-1)\varepsilon^0},\\
y&=&\frac{\varepsilon_0}{\langle\mu^2\varepsilon\rangle},\\
z&=&\frac{d+2}{d\langle\mu^2\rangle+2/y}.
\end{eqnarray}
Note that the Bruggeman equation for $\varepsilon_0$ reads 
$\langle\mu\rangle=1$, or alternatively $\langle\mu
\varepsilon\rangle=\langle\mu\rangle\varepsilon_0=\varepsilon_0$. 
From these relations, we check that: (i) $\langle\varepsilon 
E\rangle=\varepsilon^0 \langle E\rangle$; (ii) $\langle\varepsilon 
E^2\rangle=\varepsilon^0 \langle E\rangle^2$, where $\langle 
E\rangle=E^0$, as must be. Expressions for $C^\alpha_\parallel$ 
and $C^\alpha_\perp$ are readily obtained from (\ref{ee}). 

In Fig.\ 1 are displayed curves for the probability density 
${\cal P}_\parallel(E_\parallel/E^0)$ 
of the normalized longitudinal electric field, 
$E_\parallel={\bf E}\cdot \hat{\bf E}^0$, 
for a binary medium with dielectric ratio 
$\varepsilon_2/\varepsilon_1=5$, 
for various fractions $p_2$ 
of material 2. The space dimension is $d=2$. 
With ${\cal P}(E)$ given by (\ref{multimod}), and 
${\mathbf E}$ written as 
${\mathbf E}= E_\parallel\hat{\bf E}^0+{\mathbf E}_\perp$ 
(which defines ${\mathbf E}_\perp$), 
${\cal P}_\parallel$ is obtained as
\begin{equation}
{\cal P}_\parallel(E_\parallel/E^0)
=E^0\int d^{d-1}\! E_\perp\, 
{\cal P}({\mathbf E})
=E^0 \sum_\alpha p_\alpha 
\frac{1}{(2\pi C^\alpha_\parallel)^{1/2}}
e^{-\frac{1}{2} \frac{(E_\parallel-M^\alpha)^2}{C^\alpha_\parallel}},
\end{equation}
where $E_\perp$ is the transverse component of the field: 
${\mathbf{E}}_\perp={\mathbf{E}}-{\mathbf{E}}_\parallel$. 
The probability density tends towards a Dirac peak, centered 
at $E_\parallel/E^0=1$ as $p_2\to 0$ or $p_2\to 1$, since 
the field is then equal to the applied field ${\mathbf E}^0$. 
For intermediate volume fractions, the bimodal character of 
the probability distribution is evident. 

Under the same conditions, Fig.\ 2 displays the probability 
density of the scaled modulus of the transverse electric field, 
$E_\perp/E^0$. This distribution is defined, for $E_\perp>0$, 
by
\begin{equation}
{\cal P}_\perp(E_\perp/E^0)={E^0}^{d-1}\int dE_\parallel 
d\Omega_{{\mathbf E}_\perp}\, {\cal P}({\mathbf E})=
E^0 \sum_\alpha p_\alpha \frac{S_{d-1}}
{(2\pi C^\alpha_\perp)^{(d-1)/2}}\left(\frac{E_\perp}
{E^0}\right)^{d-2}e^{-\frac{1}{2} \frac{E_\perp^2}
{C^\alpha_\perp}},
\end{equation}
where $S_d=2\pi^{d/2}/\Gamma(d/2)$ is the surface of a 
$d$-dimensional unit sphere. Unlike the previous one, 
this distribution is not multimodal since the mean value 
of the transverse field is $\langle {\mathbf E}_\perp\rangle_\alpha
={\mathbf 0}$ in each phase, whatever the volume fractions. 

Fig.\ 3 displays ${\cal P}_\parallel(E_\parallel/E^0)$ for a higher 
dielectric ratio $\varepsilon_2/\varepsilon_1=1000$. 
The concentrations are $p_2=0.12$, $0.24$, $0.36$. These plots 
can be compared to that obtained from hard disks simulations by 
Cheng and Torquato (CT) in Fig.\ 3 of Ref.\ \cite{CHEN97}, for 
the same dielectric contrast and concentrations $p_2=0.2$, $0.4$, 
$0.6$. These concentrations differ from ours by a factor $0.6$. 
Apart from these differences, the overall features (shape and 
heights) of the probability distributions  are well rendered: 
a Dirac peak close to $E_\parallel=0$ indicates that the field 
is almost null in phase $2$, and a widely spread contribution 
from phase 1 for the highest $p_2$ indicates an enhancement 
of the fluctuations in this phase as the percolation 
threshold (for the Bruggeman theory), or the jamming 
threshold (for the CT simulations) is approached. The 
differences in the values for $p_2$ can be explained by 
the fact that the present theory relies on the Bruggeman 
effective-medium formula, of relevance for cell-materials 
but not really adequate for hard disks. Moreover, the 
overall agreement between our distributions and that of 
CT is not so good at moderate dielectric contrast: Bruggeman's 
theory completed by the gaussian approximation somewhat 
overestimates, because of its percolating nature, the width 
of the probability distributions. Better adequation with the 
CT simulations would probably be obtained with an 
effective-medium theory for dielectric-coated inclusions,
 which would prevent percolation. Since our main objective 
however is to discuss an effective-medium therory for WNL 
composites, such an improvement will not be considered here.

With these results in hand, analytical investigations of the 
theory can be carried out in some limiting cases, as well as 
numerical ones in more complicated situations. The next 
section examines particular WNL potentials.

\section{Weakly nonlinear power-law potential}
\label{wnpp}
As a first application, a WNL local potential of the form 
\begin{equation}
w_x(E)=\frac{1}{2}\varepsilon E^2+\frac{1}{4}\chi E^4+O(E^6), 
\end{equation}
is considered, where the $E^4$ term is the first in a series of corrections 
to the linear behaviour. The corresponding constitutive law is
\begin{equation}
D_i=[\varepsilon+\chi E^2+O(E^4)] E_i. 
\end{equation}
This approximation actually is a weak-field one.

\subsection{Effective-medium formula}
\label{scf}
The effective potential takes the form
\begin{equation}
W(E^0)=\frac{1}{2}\varepsilon^0 {E^0}^2+\frac{1}{4}\chi^0 {E^0}^4 
+O({E^0}^4),
\end{equation}
where $\varepsilon^0$ is the Bruggeman result, and \cite{STRO88,BERG89}
\begin{equation}
\chi^0=\langle \chi E^4\rangle/{E^0}^4=\sum_\alpha p_\alpha \chi_\alpha 
\langle E^4\rangle_\alpha/{E^0}^4.
\end{equation}
The phase averages $\langle E^4\rangle_\alpha$ are carried out with the 
help of Wick's theorem which allows to write down expressions for the 
integer moments of a {\em centered} vector gaussian distribution by 
mere inspection \cite{LEBE92}: with $A_i=E_i-M^\alpha_i$, 
\begin{equation}
\langle A_iA_iA_jA_j\rangle_\alpha
=\langle A_iA_i\rangle_\alpha\langle A_jA_j\rangle_\alpha
+2\langle A_iA_j\rangle_\alpha\langle A_iA_j\rangle_\alpha.
\end{equation}
We deduce
\begin{eqnarray}
\langle E^4\rangle_\alpha&=&
\langle E_i E_i E_j E_j\rangle_\alpha\nonumber\\
&=&C_{ii}^\alpha C_{jj}^\alpha
+2C_{ij}^\alpha C_{ij}^\alpha+2{M^\alpha}^2C_{ii}^\alpha
+4M^\alpha_iM^\alpha_jC_{ij}^\alpha+{M^\alpha}^4\nonumber\\
&=&\left[C_\parallel^\alpha+(d-1)C_\perp^\alpha
+{M^\alpha}^2\right]^2+2\left[{C_\parallel^\alpha}^2
+(d-1){C_\perp^\alpha}^2+2{M^\alpha}^2C_\parallel^\alpha\right].
\end{eqnarray}
A few simplifications yield the simple result
\begin{equation}
\label{csc}
\chi^0=\left[(1+2/d) y^2+(2-2/d)z^2-2\right]\langle \mu^4\chi\rangle.
\end{equation}

As a first check, we remark that in a non-disordered situation 
where $\varepsilon$ and $\chi$ are constant in the medium, we 
have $\varepsilon_0=\varepsilon$ and $\mu=1$, so that $y=z=1$; 
whence $\chi_0=\chi$, as was expected.

\subsection{Weak-contrast expansion}
A further check for the self-consistent formula (\ref{csc}) consists 
in examining its weak-contrast limit. An exact expression is known 
for any nonlinear potential, which is first briefly reminded. In 
the weak-contrast expansion \cite{BLUM89,PONT92,PONT97,BART98}, 
the local potentials $w_x(y)$ are assumed to fluctuate weakly 
around their mean value $w^{(0)}(y)=\langle w_x(y)\rangle$. 
Introducing a bookkeeping parameter $t$ to be set to 1 in the 
final results, the contrast, $w^{(1)}_x(y)$, is defined by
\begin{equation}
w_x(y)=w^{(0)}(y)+w^{(1)}_x(y)t,
\end{equation}
so that $\langle w^{(1)}_x(y)\rangle=0$. An expansion for the 
effective potential is sought for in the form:
\begin{equation}
W(E^0)=W^{(0)}+W^{(1)}t+W^{(2)}t^2+\cdots
\end{equation}
Then \cite{PONT97}, 
\begin{equation}
\label{general}
W(E^0)=\langle w(E^0)\rangle-\frac{1}{2}\frac{n_\parallel}
{\varepsilon_\parallel}\left\langle\left[\partial_i w_x^{(1)}
(E^0) {\hat E^0}_i\right]^2\right\rangle t^2+ O(t^3),
\end{equation}
where
\begin{eqnarray}
\label{depol}
n_\parallel&=&-\int \frac{d\Omega_{\hat k}}{S_d}
\frac{r ({\mathbf\hat k}\cdot{\mathbf\hat E}^0)^2}
{1+(r-1)({\mathbf\hat k}\cdot{\mathbf\hat E}^0)^2},\\
\partial_i\partial_j w^{(0)}(E^0)&\equiv&
\varepsilon_\parallel {\hat E^0}_i{\hat E^0}_j
+\varepsilon_\perp(\delta_{ij}-{\hat E^0}_i{\hat E^0}_j),\\
r&=&\varepsilon_\parallel/\varepsilon_\perp.
\end{eqnarray}
This result is the exact one \cite{PONT97}. An analogous 
expansion for the complementary potential $\tilde{W}(D^0)$ 
defined from the dual homogeneization problem can be written 
down \cite{PONT97,BART98}.

Let us therefore set $\varepsilon=\langle \varepsilon\rangle
+\delta\varepsilon\, t$ and $\chi=\langle \chi\rangle+
\delta\chi\,t$. The Bruggeman permittivity $\varepsilon^0$ 
is exact to second order in the constrast:
\begin{equation}
\varepsilon^0=\langle \varepsilon\rangle
\left[1-\frac{\langle\delta\varepsilon^2\rangle}
{d\langle\varepsilon\rangle^2}t^2+O(t^3)\right].
\end{equation}
Expanding (\ref{csc}) to second order in $t$ entails 
\cite{BART98}
\begin{equation}
\label{wnlwc}
\chi^0=\langle\chi\rangle
\left\{1+\left[\frac{2(d+8)}{d(d+2)}
\frac{\langle\delta\varepsilon^2\rangle}
{\langle\varepsilon\rangle^2}-\frac{4}{d}
\frac{\langle\delta\varepsilon\,\delta\chi\rangle}
{\langle\varepsilon\rangle\langle\chi\rangle}\right]t^2
+O(t^3)\right\},
\end{equation}
which can directly be obtained from a first-order 
expansion of (\ref{general}) in $\chi$.

Comparing now formula (\ref{csc}) to the widely used approximation
\begin{equation}
\label{chi02}
\chi^0\simeq\chi^0_2=\sum_\alpha \chi_\alpha 
\langle E^2\rangle_\alpha^2/{E^0}^4=y^2\langle \mu^4\chi\rangle, 
\end{equation}
we see that the latter is {\em not} exact to second order 
in the contrast, save in $d=1$: its expansion indeed reads
\begin{equation}
\chi^0_2=\langle\chi\rangle
\left\{1+\left[\frac{2(d+2)}{d^2}\frac{\langle\delta\varepsilon^2\rangle}
{\langle\varepsilon\rangle^2}-\frac{4}{d}
\frac{\langle\delta\varepsilon\,\delta\chi\rangle}
{\langle\varepsilon\rangle\langle\chi\rangle}\right]t^2+O(t^3)\right\}.
\end{equation}
Hence, the gaussian decoupling which accounts for the vector 
character of the electric field, of importance in nonlinear 
problems, is superior to the simple approximation $\langle 
E^4\rangle\simeq \langle E^2\rangle^2$.  

\subsection{Dilute limit}
For a binary medium with two components of constitutive 
parameters $(\varepsilon_1,\chi_1)$ and $(\varepsilon_2,\chi_2)$, 
the dilute limit is the limiting situation where the volume 
fraction $p$ of (e.g.) component 2 is small.
Setting
\begin{equation}
T=\frac{\varepsilon_2-\varepsilon_1}{\varepsilon_2+(d-1)\varepsilon_1},
\end{equation}
an expansion, to first order in $p$ of (\ref{csc}) yields
\begin{equation}
\label{myres}
\chi^0=\chi_1+\left\{(\chi_2-\chi_1) 
(T-1)^4+\chi_1T^2\left[2\frac{d(d+8)}{d+2}-4T+T^2\right]\right\}p+O(p^2).
\end{equation}
Bergman computed the exact effective nonlinear 
conductivity of a binary medium in the dilute limit, 
for spherical inclusions \cite{BERG89,ZHAN94}. His result takes the form
\begin{eqnarray}
\chi^0_{\text{exact}}&=&
\chi_1+\biggl\{(\chi_2-\chi_1) (T-1)^4\nonumber\\
\label{bergres}
&+&\chi_1T^2\left[2\frac{d(d+8)}{d+2}
+4\frac{d(d-4)}{d+2}T+\frac{d(3d^2-10d+16)}{3(d+2)}T^2\right]\biggr\}p+O(p^2).
\end{eqnarray}
Compared to (\ref{bergres}), expression (\ref{myres}) 
becomes exact when $\chi_1=0$ (nonlinear inclusion in 
a linear host). It also becomes exact for $d=1$, and 
in the limit $d\to\infty$ where the field is $E=E_0$ 
in each phase so that $\chi^0=\langle \chi\rangle$. 
Moreover, it is exact up to order $T^2$, which is 
consistent with its correct limiting weak-contrast 
behavior. For $d=2$, it is exact up to order $T^3$. 
However, the term of order $T^4$ is wrong for 
$2\le d<\infty$.
The reason for this misbehaviour is that the dilute limit requires 
an exact computation of $\langle E^4\rangle_\alpha$ (to linear order) 
\cite{BERG89}, whereas we approximate it through a gaussian average.

On the other hand, expression (\ref{chi02}) yields
\begin{equation}
\chi^0_2=\chi_1+\left\{(\chi_2-\chi_1) (T-1)^4+
\chi_1T^2\left[2(d+2)-4T+T^2\right]\right\}p+O(p^2),
\end{equation}
once again a result less precise than (\ref{myres}).

At the present time, since it requires an exact solution for 
the one-body problem, the test of the dilute limit appears to be 
the most challenging one for nonlinear effective-medium theories. 

This test is illustrated in Fig.\ 4 which displays $\chi^0$, as 
calculated from (\ref{csc}) and (\ref{chi02}), against the 
concentration $p$ of medium 2, for moderate dielectric contrast 
and $d=2$. The thick line segments at $p=0,1$ represent exact 
tangents obtained from (\ref{bergres}). The tangent at $p=1$ 
follows from substituting $(1-p)$ for $p$ and interchanging the 
indices $1$ and $2$ in (\ref{bergres}). Though not exact in the 
dilute limit, formula (\ref{csc}) yields tangents quite close to 
the exact ones. The marked inaccuracy of formula (\ref{chi02}) 
near $p=0$ and $p=1$ results in a lower value for $\chi^0$ in 
the whole concentration range.  

\subsection{Percolative behavior}
Before examining the predictions of Equ.\ (\ref{csc}) near the 
percolation transition, I first briefly review the critical 
behaviour of WNL composites, and discuss a flaw of the 
Bruggeman formula in this context.
 
An insulator/perfect conductor binary mixture undergoes a 
percolation transition for a critical metal fraction 
$p=p_c$ \cite{CLER90}. For WNL phases, the critical behaviour 
of the field fluctuations is now well understood 
\cite{WRIG86}. In the limiting situation where 
$\varepsilon_1\ll \varepsilon_2$, and for $p\leq p_c$, 
one observes a behaviour
\begin{equation}
\varepsilon^0\propto \varepsilon_1(p-p_c)^{-s},
\qquad\chi^0\propto \chi_1(p-p_c)^{-(2s+\kappa')};
\end{equation}
on the contrary, for $p>p_c$, one has
\begin{equation}
\varepsilon^0\propto \varepsilon_2(p_c-p)^{t},\qquad\chi^0
\propto \chi_2(p_c-p)^{(2t-\kappa)},
\end{equation}
where $s$ and $t$ are the superconductivity and conductivity 
exponents, and where $\kappa$ and $\kappa'$ are the noise 
exponents which characterize an anomalous nonlinear 
susceptibility enhancement near the threshold 
\cite{RAMM85,WRIG86}. As a consequence of the exact 
inequality $\langle E^4\rangle\ge\langle E^2\rangle^2$, 
the exponents $\kappa$ and $\kappa'$ are necessarily 
positive. 

In the Bruggeman formula, $p_c=1/d$, and $\varepsilon_0
\simeq\varepsilon_1/(1-dp)$ for $p<1/d$, $\varepsilon_0
\simeq\varepsilon_2(1-dp)/(1-d)$ for $p>1/d$ 
($\varepsilon_1\ll \varepsilon_2$), whence $s=t=1$. 
Reporting these expressions into (\ref{csc}) yields 
$\kappa=\kappa'=0$, a result shared by all 
effective-medium formulae based on (\ref{derivscal}). 

These values can be compared to exact bounds obtained 
for $\kappa$ and $\kappa'$ on the basis of the 
links-nodes-blob (LNB) model for electric percolation, 
now commonly accepted as a model for e.g.\ random 
resistor networks (RRNs). These are \cite{WRIG86}
\begin{mathletters}
\label{bounds}
\begin{eqnarray}
&&(3d-4)\nu-2t+1\le \kappa\le 2(d-1)\nu-t,\\
&&(4-d)\nu-2s+1\le\kappa'\le 2\nu-s,
\end{eqnarray}
\end{mathletters}
(for $d\le 6$; the values for $d>6$ are that of $d=6$. 
Moreover $t$ is not defined for $d=1$), where $\nu$ is 
the correlation length exponent: $\xi\propto |p-p_c|^{-\nu}$. 
Though the latter information is absent from the Bruggeman 
theory, an inequality $\nu>0$ must hold in any situation 
relevant to percolating systems. In (\ref{bounds}), the 
lower bounds must not be greater than the upper ones, 
which leads to lower bounds for the exponents $s$ and 
$t$ themselves, namely
\begin{eqnarray}
&&s\ge (2-d)\nu+1,\\
&&t\ge (d-2)\nu+1.
\end{eqnarray}
In the Bruggeman theory, $s=t=1$ so that only $\nu=0$ 
-- a problematic value by itself -- is allowed, save for 
$d=2$. Reporting $\nu=0$ into (\ref{bounds}) leads to 
$\kappa=\kappa'=-1$, which is unacceptable. For $d=2$, 
one obtains $\kappa=\kappa'=2\nu-1$. Compared to the 
values $\kappa=\kappa'=0$ of the above effective-medium 
theory, this gives $\nu=1/2$, a non-conflicting value, 
though lower than the exact result $4/3$. 

Of course, the fact that Bruggeman's formula is a poor 
approximation near the percolation threshold is well-known. 
However, this discussion enlightens a fundamental 
inconsistency in the Bruggeman exponent values which 
goes beyond mere numerical inadequacy, as long as one
 wishes to estimate the properties of systems obeying 
the LNB scheme. It is to be noted that for $d=2$, where
 no inconsistency appears, the exact equality $s=t$ holds 
(a consequence of self-duality) \cite{STRA77} and is obeyed 
by the Bruggeman exponents. 

The above argument is to be brought together with another one 
by Bergman. Discussing the failure of a ``non ambiguous'' 
non-linear Bruggeman-type model, he concluded that bounds 
for fluctuations always refer to some particular type of 
microstructure, and that no such bounds exist which apply 
to all materials \cite{BERG89}.

Because of these problems, comparisons between 
effective-medium theories built on the Bruggeman 
formula, and simulations on RRNs are expected to 
be more significant in dimension $d=2$ (another 
reason for preferring the two-dimensional case 
as a test-bench is that the bond percolation 
threshold $p_c=1/2$ on a square lattice is exactly 
reproduced by Bruggeman's formula). In Fig.\ 5 are 
displayed formulas (\ref{csc}) and (\ref{chi02}) 
against $p$, in a high contrast situation  
($\varepsilon_2/\varepsilon_1=10000$). The same 
trend as in Fig.\ 4 is observed: formula (\ref{csc}) 
yields a higher estimate. The values for 
$\varepsilon_{1,2}$ and $\chi_{1,2}$ are the 
same as that used in Fig.\ 3 of a paper by Levy 
and Bergman \cite{LEVY94}, where simulations results 
on RRNs are reported, and compared to (\ref{chi02}). 
The authors remarked that the height of the peak in 
$\chi^0$ at $p_c$ was badly underestimated by (\ref{chi02}), 
which gives $\chi^0_{\text{peak}}\simeq 0.5$ only. In 
their numerical simulations, for a $30\times 30$ network, 
the peak height is found to be 
$\chi^0_{\text{peak}}\simeq 0.86$. In the 
infinite-size limit, this value is expected 
to increase somewhat. In Fig.\ 5, the height 
of the peak given by (\ref{csc}) is 
$\chi^0_{\text{peak}}\simeq 1.0$, which thus 
compares fairly well to simulations.

\subsection{Other power-law nonlinearities}
In this section, potentials of the type (\ref{orig}) 
where \cite{AHAR87}
\begin{equation}
w^{\text{nl}}_x(E)=\frac{\chi(x)}{\gamma+1}E^{\gamma+1}
\end{equation}
are considered. For simplicity, $\gamma$ is assumed to be 
constant in the material. The gaussian averages over the 
field in each phase are carried out numerically, with the 
help of a bi-variate integration routine. One integration 
variable is the field modulus $E$, and the other is the 
cosine ${\mathbf E}\cdot{\bf E}^0/(E E^0)$. In fig.\ 
\ref{fig6} are shown plots for \begin{equation}
\label{directg}
\chi^0=\langle\chi (E/E^0)^{\gamma+1}\rangle,
\end{equation}
computed from gaussian averages, for a moderate 
dielectric contrast and various powers $\gamma>1$ 
ranging from $\gamma=1.5$ to $\gamma=5$, against 
the volume fraction $p$ of medium 2 (solid lines). 
An exponential enhancement of $\chi^0$ is observed 
with increasing $\gamma$. Its origin lies in the 
existence of a non-zero probability for $E>E^0$ 
in the composite, since ${\bf E}^0=\langle{\bf E}\rangle$ 
by definition. These fluctuations are amplified by the 
nonlinearity, while that for which $E<E^0$ are reduced. 
The larger the width of the probability distribution, 
the larger this enhancement, so that the peak 
culminates in the region $p\simeq p_c=1/2$. Once 
again, the estimates of the present theory are much 
higher than that predicted by the decoupling assumption 
\begin{equation}
\label{decouplg}
\chi^0_2=\sum_\alpha p_\alpha \chi_\alpha \langle(E/E^0)^{2}
\rangle_\alpha^{(\gamma+1)/2},
\end{equation}
(dashed lines), especially for large values of $\gamma$. 
This is understandable, since the functions to be averaged can 
be written $w_x^{\text{nl}}(E)=\chi(x)f(E^2)$, where 
$f(z)=z^{(1+\gamma)/2}/(1+\gamma)$. For $\gamma>1$, these functions 
are convex, so that $\langle f(z)\rangle\geq f(\langle z\rangle)$.
 Therefore, the decoupling assumption always underestimates the 
fluctuations when applied to such potentials. An overestimation 
would instead take place with concave potentials.

\section{Conclusion}
\label{c}
In this article, a theory for the nonlinear susceptibility of 
weakly nonlinear composites is proposed. This theory is based 
on a multimodal gaussian approximation for the overall 
probability distribution of the electric field, to linear 
order. The parameters which define this distribution (means 
and second moments, in each phase of the disordered medium), 
are obtained from Bruggeman's theory, whose limitations are 
discussed in this context. The present model provides for the 
first time an analytical estimate for the probability distribution 
of the electric field, for arbitrarily high dielectric contrast, 
in percolating media. The resulting effective nonlinear susceptibility 
is exact to second order in the contrast, and close to the exact result 
in the dilute limit. Significant quantitative improvement is obtained 
on previous nonlinear effective-medium theories, even in the percolating 
regime, at least in the two-dimensional case. This study emphasizes the 
importance of accounting for the vector character of the electric field
 when averaging the local potentials. Improvements of the present theory 
could consist in finding a better approximation than the gaussian one 
for the components of the probability distribution of the field. More 
realistic distributions could be obtained by, e.g., extending the 
perturbative theory of Barth\'el\'emy and Orland \cite{BART97} to high 
dielectric contrasts via self-consistent effective-medium approximations.


\acknowledgements
Stimulating discussions with M.\ Barth\'el\'emy and H.\ Orland are gratefully 
acknowledged. I also thank H.\ E.\ Stanley and M.\ Barth\'el\'emy for their 
kind hospitality at the Center for Polymer Studies (Boston University), 
where part of this work was performed. 
\appendix


\section{Self-consistent anisotropic linear theory with permanent polarization}
\label{sltwpp}

With local potentials of the form (\ref{ferroel}), the homogeneized 
potential reads
\begin{equation}
\label{wlin}
\tilde{W}_{\text{lin}}(E^0)=\frac{1}{2}E^0_i \varepsilon^0_{ij}E^0_j
+P^0_i E^0_j +\frac{1}{2}\langle \Delta P_i \mu_{ij} g_{jk} \Delta P_j\rangle,
\end{equation}
where
\begin{eqnarray}
g_{ij}&=&-\int \frac{d\Omega_{\hat {\bf k}}}{S_d} \frac{ k_i k_j}
{k_l \varepsilon^0_{lm} k_m},\\
\mu_{ij}^\alpha&=&[1-g(\varepsilon^\alpha-\varepsilon^0)]^{-1}_{ij},\\
P^0_i&=&\langle P_j\mu_{ji}\rangle,\\
\Delta P_i^\alpha&=&P_i^\alpha-P^0_i,
\end{eqnarray}
and the Bruggeman condition for $\varepsilon^0_{ij}$ is 
$\langle \mu_{ij}\rangle=\delta_{ij}$.

Formula (\ref{wlin}) has been derived by means of a functional formalism 
\cite{PELL99}, used with a trial potential of the form \cite{PELL99b} 
$W^0(E)=(1/2)E_i\varepsilon^0_{ij} E_j+P^0_i E_i+c^0$. Rather than to 
give a demonstration which would complicate the present article, the 
validity of (\ref{wlin}) is established by looking at its consequences. 
First, one can easily check that, for a binary medium, (\ref{wlin}) 
exactly reduces to formula (2.16) in the work by Ponte Casta\~neda 
and Kailasam \cite{PONT97}. Next, the average field per phase, 
$\langle E_i\rangle_\alpha=M^\alpha_i$, is (using the symmetry of the 
tensors $g$ and $\mu g$; $\mu$ itself is not necessarily symmetric)
\begin{equation}
\label{eav}
M^\alpha_i=\frac{1}{p_\alpha}\frac{\partial 
\tilde{W}_{\text{lin}}(E^0)}{\partial P_i^\alpha}=\mu^\alpha_{ij}
( E^0_j+g_{jk}\Delta P_k^\alpha).
\end{equation}
From the definition of $P^0_i$, one deduces that $\langle M\rangle=E^0$, 
in agreement with the boundary conditions. Finally, the macroscopic 
consitutive relation derived from the effective potential, namely
\begin{equation}
D^0_i=\frac{\partial \tilde{W}_{\text{lin}}(E^0)}{\partial E^0_i}
=\varepsilon_{ij}^0 E^0_j+P^0_i,
\end{equation}
is consistent with the equation
\begin{equation}
D^0_i=\langle \varepsilon_{ij} E_j+P_i\rangle
=\langle\varepsilon^\alpha_{ij} M_j^\alpha +P_i^\alpha\rangle,
\end{equation}
with $M_j^\alpha$ computed with (\ref{eav}). This directly follows 
from the equivalent form of the Bruggeman equation, 
$\langle \varepsilon_{ij}\mu_{jk} \rangle=\varepsilon^0_{ij}\langle 
\mu_{jk}\rangle=\varepsilon^0_{ij}$, and from the identity 
$\mu_{ij}^\alpha=\delta_{ij}+\mu_{ik}^\alpha g_{kl}
(\varepsilon_{li}^\alpha-\varepsilon^0_{li})$.



\twocolumn
\begin{figure}
\narrowtext
\vspace*{0.0cm}
\centerline{
\epsfysize=0.9\columnwidth{\rotate[r]{\epsfbox{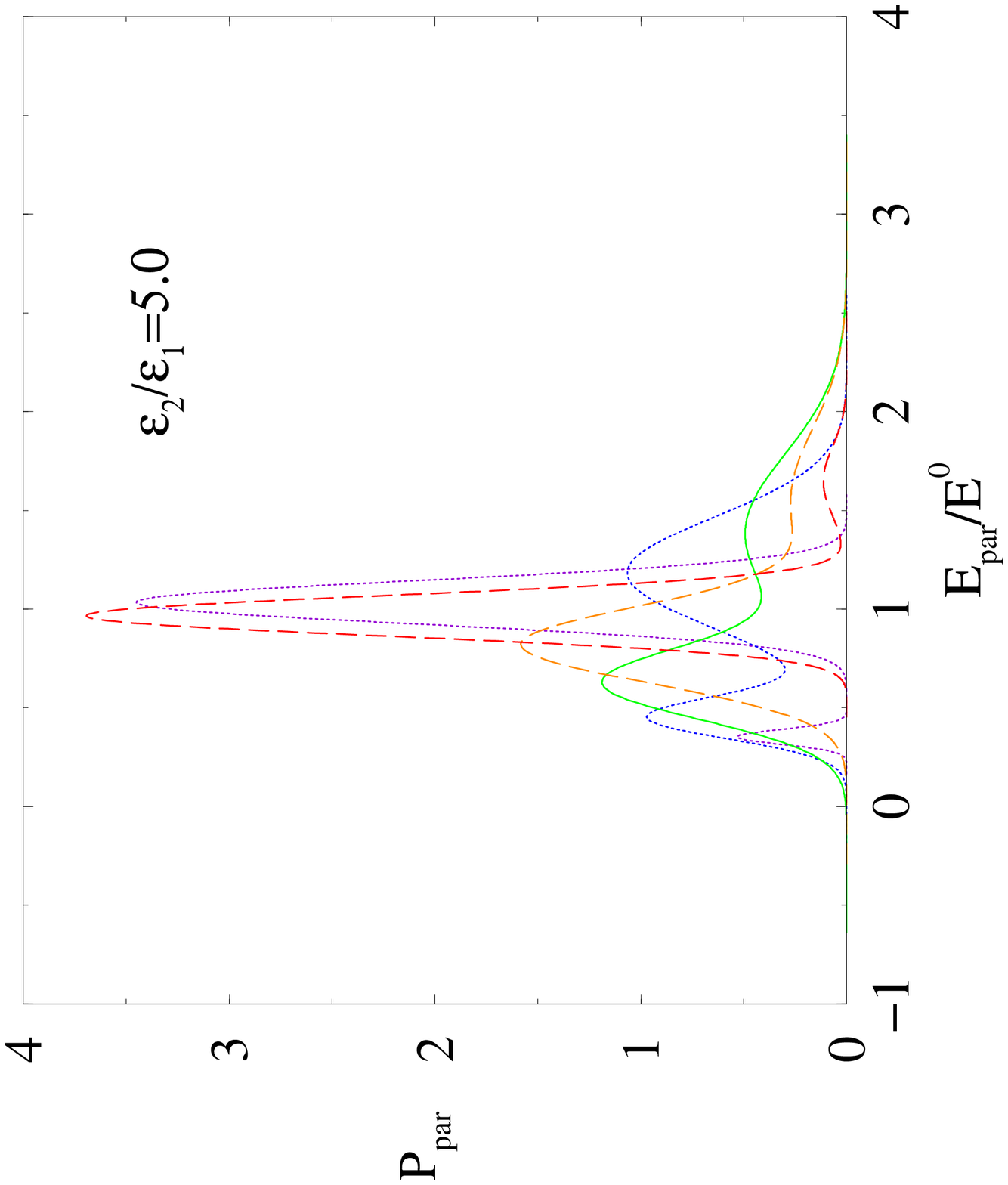}}}}
\vspace*{0.0cm}
\caption{Field probability density function 
${\cal P}_\parallel(E_\parallel/E^0)$ 
($E_\parallel={\bf E}\cdot\hat{\bf E}^0$) in the 
Bruggeman approximation, for various volume fractions 
$p_2$ of component 2. The dielectric constant ratio is 
$\varepsilon_2/\varepsilon_1=5$. Dots: $p_2=0.05$ 
(violet,highest curve); $p_2=0.25$ (blue,lowest curve). 
Solid line: $p_2=0.5$ (green). Dashes: $p_2=0.75$ 
(orange, lowest curve); $p_2=0.95$ (red, highest curve).
}
\label{fig1}
\end{figure}

\begin{figure}
\narrowtext
\vspace*{0.0cm}
\centerline{
\epsfysize=0.9\columnwidth{\rotate[r]{\epsfbox{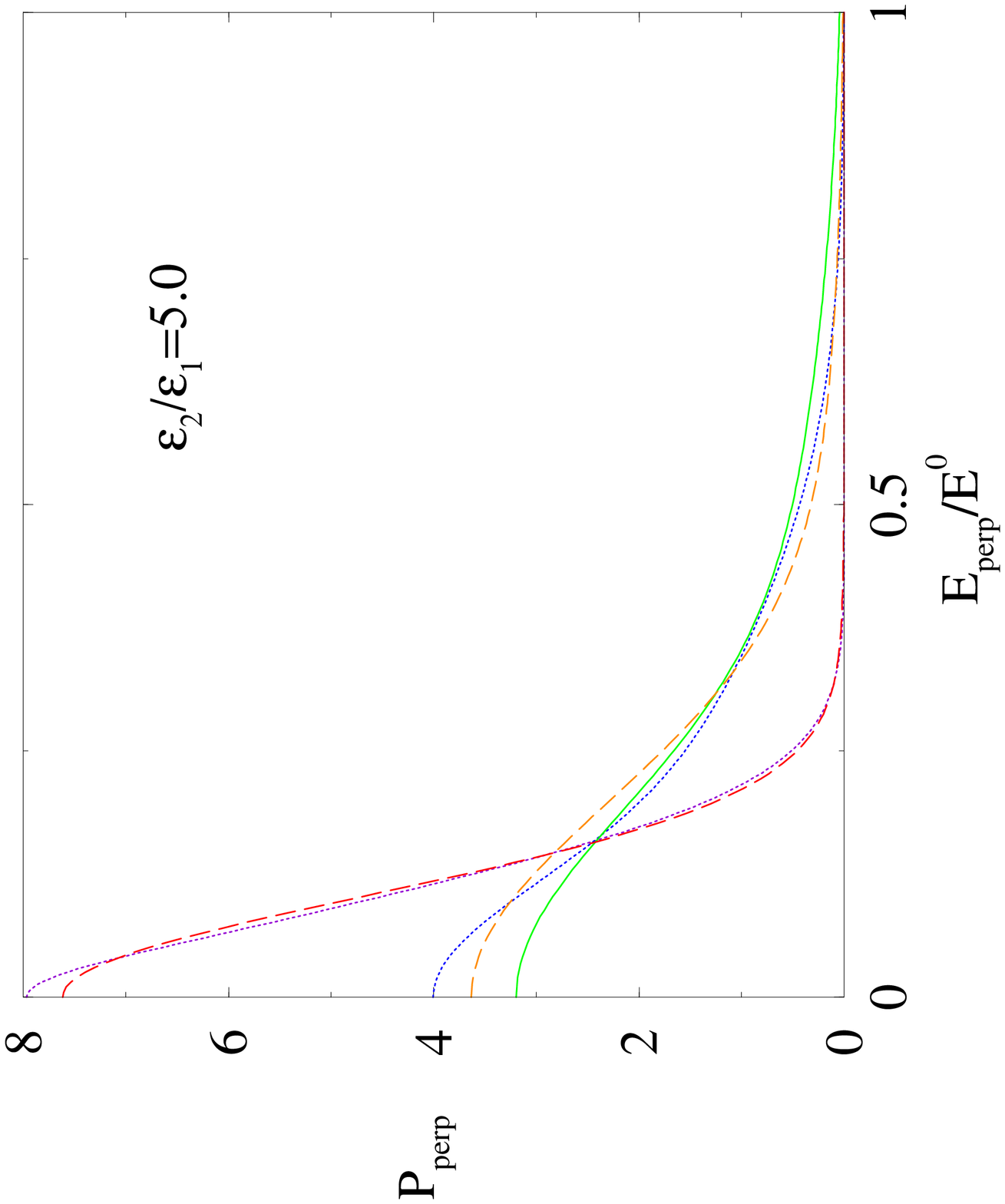}}}}
\vspace*{0.0cm}
\caption{Field probability density function 
${\cal P}_\perp(E_\perp/E^0)$ 
($E_\perp=|\!|{\bf E}-E_\parallel{\bf E}^0|\!|$) 
in the Bruggeman approximation, for various volume 
fractions $p_2$ of component 2. The dielectric 
constant ratio is $\varepsilon_2/\varepsilon_1=5$. 
Dots: $p_2=0.05$ (violet, highest curve); $p_2=0.25$ 
(blue, lowest curve). Solid line: $p_2=0.5$ (green). Dashes: 
$p_2=0.75$ (orange, lowest curve); $p_2=0.95$ (red, highest curve).
}
\label{fig2}
\end{figure}

\begin{figure}
\narrowtext
\vspace*{0.0cm}
\centerline{
\epsfysize=0.9\columnwidth{\rotate[r]{\epsfbox{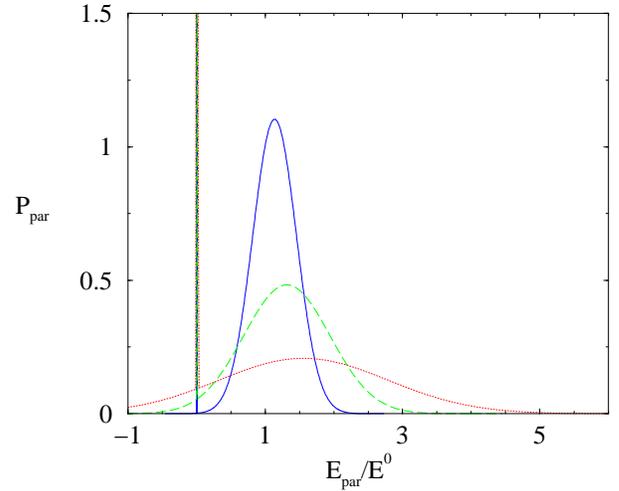}}}}
\vspace*{0.0cm}
\caption{Field probability density function 
${\cal P}_\parallel(E_\parallel/E^0)$ 
($E_\parallel={\bf E}\cdot\hat{\bf E}^0$) in 
the Bruggeman approximation, for various volume 
fractions $p_2$ of component 2. The dielectric 
constant ratio is $\varepsilon_2/\varepsilon_1=1000$. 
Solid: $p_2=0.12$; dashes: $p_2=0.24$; dots: $p_2=0.36$.\\
\\
\\
}
\label{fig3}
\end{figure}

\begin{figure}
\narrowtext
\vspace*{0.0cm}
\centerline{
\epsfysize=0.9\columnwidth{\rotate[r]{\epsfbox{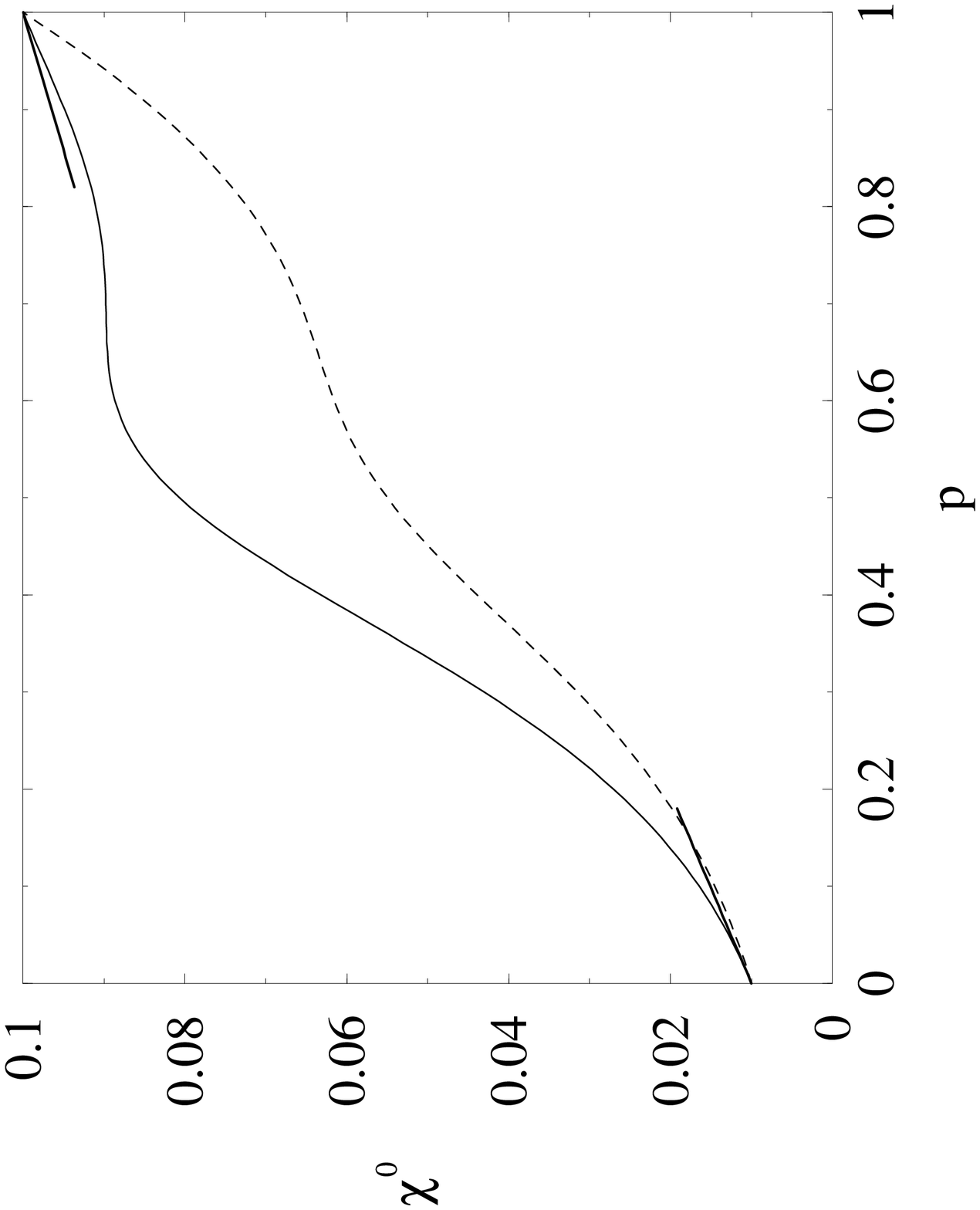}}}}
\vspace*{0.3cm}
\caption{Comparison between Eqs.\ (\ref{csc}) 
(solid line) and (\ref{chi02}) (dashes). Effective 
nonlinear susceptibility vs.\ volume fraction $p$ of 
component 2. The space dimension is $d=2$. The 
parameters are: $\varepsilon_1=1$, $\varepsilon_2=10$, 
$\chi_1=0.01$, $\chi_2=0.1$. The segments at $p=0$ and 
$p=1$ represent exact slopes computed with expression 
(\ref{bergres}).}
\label{fig4}
\end{figure}

\begin{figure}
\narrowtext
\vspace*{0.0cm}
\centerline{
\epsfysize=0.9\columnwidth{\rotate[r]{\epsfbox{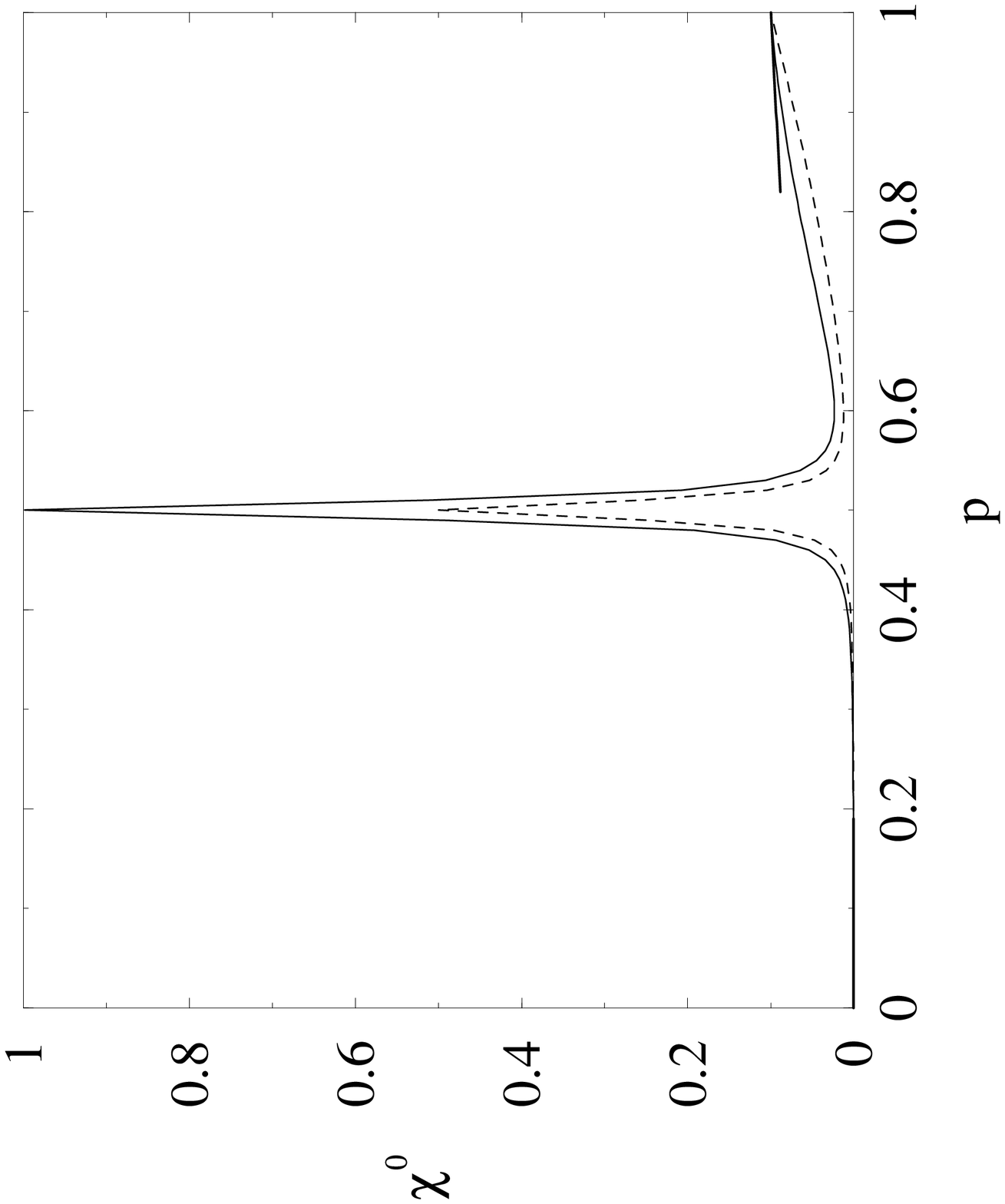}}}}
\vspace*{0.0cm}
\caption{Comparison between Eqs.\ (\ref{csc}) 
(solid line) and (\ref{chi02}) (dashes). Effective 
nonlinear susceptibility vs.\ volume fraction $p$ of 
component 2. The space dimension is $d=2$. The parameters 
are: $\varepsilon_1=1$, $\varepsilon_2=10^4$, $\chi_1=10^{-4}$, 
$\chi_2=0.1$. The segment at $p=1$ represents the exact slope 
computed with expression (\ref{bergres}).
}
\label{fig5}
\end{figure}

\begin{figure}
\narrowtext
\vspace*{0.0cm}
\centerline{
\epsfysize=0.9\columnwidth{\rotate[r]{\epsfbox{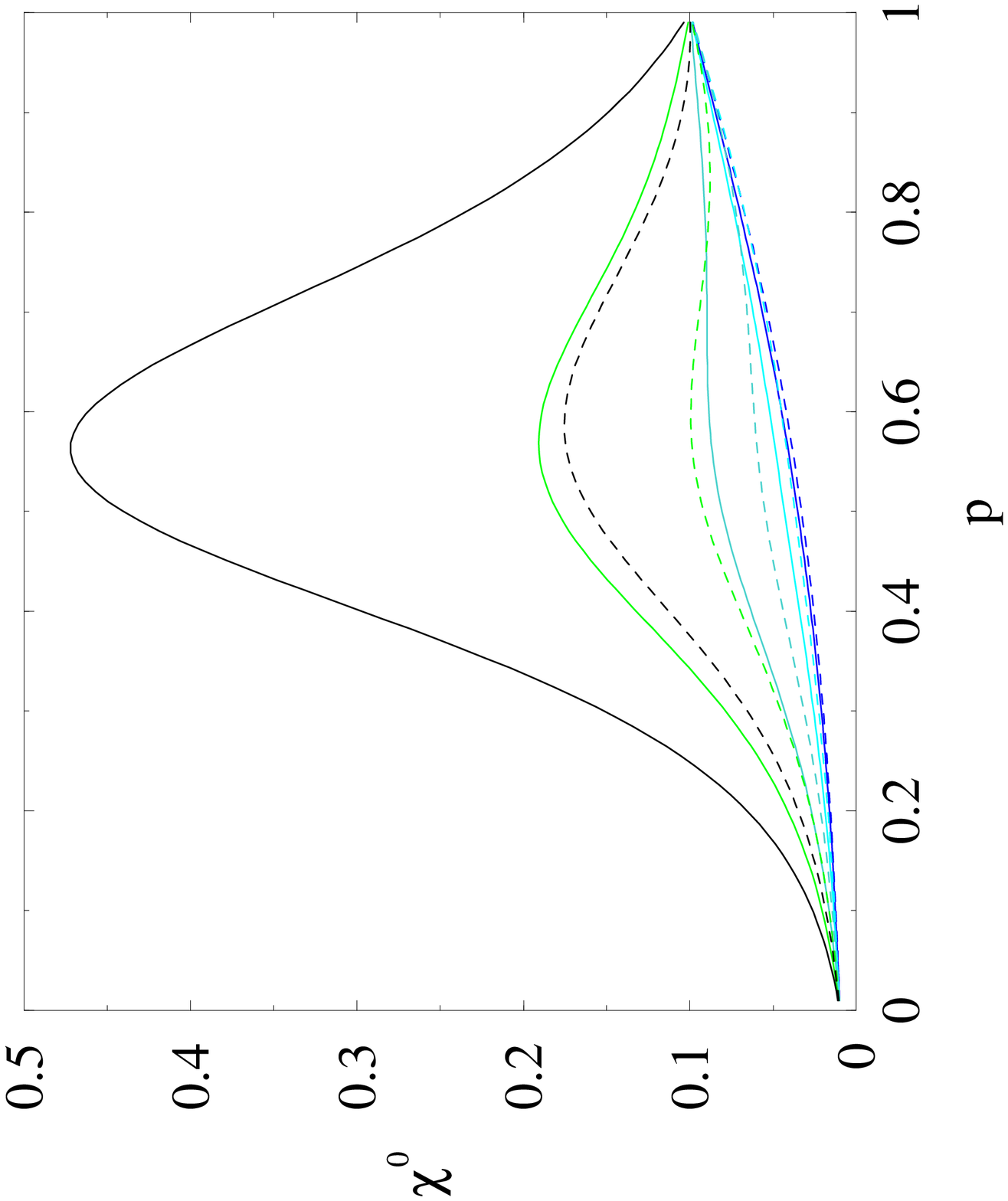}}}}
\vspace*{0.3cm}
\caption{Comparison between Eqs.\ (\ref{directg}) 
(solid line) and (\ref{decouplg}) (dashes) for power-law 
nonlinearities $w^{\text{nl}}_x=\chi(x) E^{\gamma+1}/(\gamma+1)$. 
Effective nonlinear susceptibility vs.\ volume fraction $p$ of 
component 2. The space dimension is $d=2$. The parameters are: 
$\varepsilon_1=1$, $\varepsilon_2=10$, $\chi_1=0.01$, 
$\chi_2=0.1$. From bottom to top: $\gamma=1.5$ (blue), $2$ (cyan), 
$3$ (turquoise), $4$ (green), $5$ (black).
}
\label{fig6}
\end{figure}


\begin{thebibliography}{99}
\bibitem{NOTE1} It can be shown that the boundary condition 
$\langle{E}\rangle=E^0$ truly corresponds to the more 
standard condition of uniform normal induction on the 
boundary $\partial V$, namely $n_i D_i=n_i D^0_i$ 
(see, e.g.\ Ponte-Casta\~neda \cite{PONT92b}); we then 
have $\langle{D}\rangle=D^0$; but the relationship 
(\ref{constithom}) between $D_0$ and $E_0$ is known 
only after the homogeneization problem has been solved.
\end{thebibliography}
\end{document}